\documentclass[12pt,prc,showpacs,preprintnumbers,unsortedaddress,amsmath,amssymb,floatfix]{revtex4}
\input epsf.sty
\usepackage{graphicx}
\usepackage{bm}

\begin{document}

\title{Low-lying Dipole Modes in $^{26,28}$Ne in the Quasiparticle
Relativistic Random Phase Approximation}

\author{Li-Gang Cao$^{a}$ and Zhong-Yu Ma$^{b,c}$\footnote{also Center of Theoretical
Nuclear Physics, National Laboratory of Heavy Ion Accelerator of
Lanzhou, Lanzhou 730000 and Institute of Theoretical Physics,
Chinese Academy of Sciences, Beijing 100080}}

\affiliation{$^{a}$Institute of High Energy Physics, Chinese
Academy of Sciences, Beijing 100039, P.R. of China}

\affiliation{$^{b}$ China Center of Advanced Science and
Technology (World Laboratory), P.O.Box 8730, Beijing 100080, P.R.
of China}

\affiliation{$^{c}$China Institute of Atomic Energy, Beijing
102413, P.R. of China}

\date{\today}

\begin{abstract}

The low-lying isovector dipole strengths in neutron rich nuclei $^{26}$Ne
and $^{28}$Ne are investigated in the quasiparticle relativistic random
phase approximation. Nuclear ground state properties are calculated in an
extended relativistic mean-field theory plus BCS method where the
contribution of the resonant continuum to pairing correlations is properly
treated. Numerical calculations are tested in the case of isovector dipole
and isoscalar quadrupole modes in the neutron rich nucleus $^{22}$O. It is
found that in present calculation low-lying isovector dipole strengths at
$E_x < 10$ MeV in nuclei $^{26}$Ne and $^{28}$Ne exhaust about 4.9\% and
5.8\% of the Thomas-Reiche-Kuhn dipole sum rule, respectively. The centroid
energy of the low-lying dipole excitation is located at 8.3 MeV in $^{26}$Ne
and 7.9 MeV in $^{28}$Ne.

\end{abstract}



\maketitle

\section{introduction}

Nuclear giant resonances have been known since 50 years ago for the dipole
mode and  more than 30 years for the other modes. But the research field was
limited to the excitations of nuclei along the $\beta$-stability
line\cite{Ber75,Spe91,Pit71}. Recently, the radioactive ion beam physics has
become one of the frontiers in nuclear physics. It offers the possibility to
broaden the study of the giant resonance to weakly bound nuclei. Nuclei
close to the drip line present some unique properties: a small separation
energy of the valence nucleon, the smearing density distribution and a
strong coupling between the bound state and the particle continuum. Those
exotic properties attract more attentions both experimentally and
theoretically. Low-lying electric dipole modes may appear in these weakly
bound nuclei, which are so-called Pygmy Dipole Resonances. Although carrying
only a small fraction of the full dipole strength these states are of
particular interest because they are expected to reflect the motion of the
neutron skin against the core formed with an equal number of protons and
neutrons. Recent experiments have shown that the increase of the dipole
strength at low energies in neutron-rich nuclei could affect the
corresponding radiative neutron capture cross section
considerably\cite{Gor02}, which has a significance in astrophysics. Over the
last decade, much experimental and theoretical efforts have been dedicated
to investigate properties of the low-lying dipole mode in light neutron-rich
nuclei, in particular, to answer whether or not these dipole excitations can
be attributed to the
collectivity\cite{Try03,Rye02,Lei01,Iwa00,Iwa001,Aum99,Sac93,Nak94,Aum96,Vre01,Sag01,Sag99}.

Recently, Beaumel et al.\cite{Bea} have measured the inelastic scattering of
$^{26}$Ne + $^{208}$Pb using a 60 MeV/u $^{26}$Ne secondary beam at RIKEN.
This reaction is dominated by Coulomb excitations and selective for $E1$
transitions. The experimental data are now under analysis\cite{Gib04}. As a
subsequent work, they will continue the experiment using a more neutron rich
projectile $^{28}$Ne. Therefore the theoretical investigation of low-lying
dipole modes in $^{26}$Ne and $^{28}$Ne has a practical significance.
$^{26}$Ne and $^{28}$Ne are neutron rich nuclei, whose Fermi surfaces are
close to the particle continuum. Therefore the description of those nuclei
has to explicitly include the coupling between bound states and the particle
continuum. The contribution of the particle continuum to the nuclear
properties at low-energies can mainly be attributed to a few resonant
states\cite{San00,Kru01,Gra01,Cao031,San03}. On the other hand, it is well
known that pairing correlations play an important role in describing
properties of open shell nuclei. In order to depict the collective
excitations of those nuclei pairing correlations have to be taken into
account. Recently, a number of theoretical works has devoted to study the
properties of low-lying dipole modes in the framework of the quasiparticle
random phase approximation (QRPA)\cite{Kam98,Kha00,Mat01,Hag01,Gor02}. Paar
and his co-workers\cite{Pa03} have studied the evolution of the low-lying
dipole strength in Sn-isotopes in the quasiparticle relativistic random
phase approximation (QRRPA) in the configuration space formalism.

In this paper, we aim at the investigation of the properties of
low-lying dipole modes in neutron rich nuclei $^{26}$Ne and
$^{28}$Ne in the QRRPA which is formulated in the response
function method. The QRRPA is an extension of the fully consistent
RRPA \cite{Ma01,Ring01,Ma02} by taking into account the effect of
pairing correlations. A consistent treatment of RRPA within the
RMF approximation requires the configurations including not only
the pairs formed from the occupied Fermi states and unoccupied
states but also the pairs formed from the Dirac states and
occupied Fermi states. It has been emphasized in
Refs.\cite{Ma01,Ring01} that the inclusion of configurations built
from the positive energy states in the Fermi sea and negative
energy states in the Dirac sea is essential to give correctly a
quantitative description in the excitation energies of isoscalar
giant multipole resonances as well as to ensure the current
conservation and decouple the spurious states. In present
calculations, we pay more attentions to the energy weighted moment
$m_1$ and the centroid energy of the low-lying dipole strength as
well as the contribution of states around the Fermi surface to the
low-lying dipole strength. Although some theoretical
investigations show that $^{26}$Ne and $^{28}$Ne are deformed and
strongly anharmonic\cite{Rei99,Uts99,Aza02}, a spherical symmetry
is assumed in the present investigation. In order to show the
applicability of the method with a spherical assumption we also
study the quadrupole excitations in these nuclei and compare the
calculated position and transition strength of the lowest 2$^+$
states with the experimental data.

In this work, the ground state properties of neutron rich nuclei
$^{26}$Ne and $^{28}$Ne are calculated in the extended
relativistic mean field and Bardeen-Cooper-Schrieffer (RMF+BCS)
approxiamtion\cite{Cao031}, where the resonant continuum is
properly treated. The empirical pairing gaps deduced from odd-even
mass differences are adopted in the BCS calculation in this work.
All calculations are performed with the parameter set
NL3\cite{Lal97}, which gives a good description of not only the
ground state properties\cite{Lal99} but also the collective giant
resonance\cite{Ma01,Ring01,Ma02,Vre00,Cao02,Cao03}.

The paper is arranged as follows. In Sec.II we present the formalism of the
QRRPA in the extended RMF+BCS ground state in the response function
approach. A test of the numerical calculation of the QRRPA in the neutron
rich nucleus $^{22}$O is performed and compared with the results in
Ref\cite{Pa03}, which is given in Sec.III. In Sec.IV the ground state
properties of $^{26}$Ne and $^{28}$Ne are studied in the extended RMF+BCS
approach. Then the QRRPA in the response function formalism is applied to
study the properties of isovector giant dipole resonances in nuclei
$^{26}$Ne and $^{28}$Ne. Finally we give a brief summary in Sec.V.

\section{The quasiparticle relativistic random phase approximation}

There are usually two methods to obtain the RPA strength in the
study of nuclear collective excitations. One is working in a
particle-hole configuration space and solving the RPA equation by
a matrix diagonalization method\cite{Ring01}, and another is based
on the linear response theory\cite{Ma97}. In the response function
formalism one solves a Bethe-Salpeter equation by inversion. In
both methods the starting point is a self-consistent solution of
the nuclear ground state. In this paper we shall work in the
response function formalism and study nuclear dipole excitations
in neutron rich nuclei.

In the RRPA calculation we first solve the Dirac equation and
equations of meson fields self-consistently in the coordinate
space. The continuum is discretized by expanding nucleon spinors
on a complete set basis, such as eigenfunctions in a spherical
harmonic oscillator potential. Those single particle states are
used to build the RRPA configurations: a set of particle-hole
pairs($ph$) and pairs formed from the negative energy state in the
Dirac sea and the hole state in the Fermi
sea($\overline{\alpha}h$).

The response function of a quantum system to an external field is
given by the imaginary part of the polarization operator:
\begin{equation}
R(Q,Q;{\bf k},{\bf k^{\prime }};E)=\frac 1\pi Im\Pi ^R(Q,Q;{\bf k},{\bf %
k^{\prime }};E)~,   \label{eq1}
\end{equation}
where $Q$ is an external field operator. The RRPA polarization
operator is obtained by solving the Bethe-Salpeter equation:

\begin{widetext}
\vglue -0.50cm
\begin{eqnarray}
&\Pi(Q,Q;{\bf k},{\bf k^{\prime }},E)=&\Pi _0(Q,Q;{\bf k},{\bf
k^{\prime }} ,E)-\sum_ig_i^2\int d^3k_1d^3k_2\Pi _0(Q,\Gamma
^i;{\bf k},{\bf k}_1,E) \nonumber\\
&&D_i({\bf k}_1,{\bf k}_2,E)\Pi (\Gamma _i,Q;{\bf k}_2,{\bf
k^{\prime }},E)~, \label{eq2}
\end{eqnarray}
\end{widetext}

In the RRPA, the residual particle-hole interactions are obtained from the
same Lagrangian as in the description of the nuclear ground state. They are
generated by exchanging various mesons: the isoscalar scalar meson $\sigma$,
the isoscalar vector meson $\omega$ and the isovector vector meson $\rho$.
Therefore, in Eq.(2) the sum $i$ runs over $\sigma$, $\omega $ and $\rho $
mesons, and $g_i $ and $D_i$ are the corresponding coupling constants and
meson propagators. They are $\Gamma ^i=1$ for $\sigma $ meson and $\Gamma
^i=\gamma ^\mu ,\gamma ^\mu \tau _3$ for $\omega $ and $\rho $ mesons,
respectively. The meson propagators in the non-linear model are non-local in
the momentum space, and therefore have to be calculated numerically. The
detailed expressions of non-linear meson propagators $D_i({\bf k}_1,{\bf
k}_2,E)$ can be found in Ref.\cite{Ma97}. $\Pi _0$ is the unperturbed
polarization operator, which in a spectral representation has the following
retarded form:
\begin{widetext}
\vglue -0.50cm
\begin{eqnarray}
&&\Pi _0^R(P,Q;k,k^{\prime };E) \nonumber \\
&&=\frac{(4\pi)^{2}}{2L+1}\left\{
\sum_{h,p}(-1)^{j_{h}+j_{p}}\left[ \frac{%
\langle\overline{\phi}_h\|P_L\|\phi _p\rangle\langle\overline{\phi
}_p\|Q_L\|\phi_h\rangle}{E-(\varepsilon_p-\varepsilon_h)+i\eta }-\frac{\langle\overline{\phi }_p\|P_L\|\phi _h\rangle%
\langle\overline{\phi }_h\|Q_L\|\phi _p\rangle}{E+(\varepsilon_p-\varepsilon_h)+i\eta }%
\right]\right. \nonumber \\
&&\left.~~~~~~~~~~~+\sum_{h,\overline{\alpha} }(-1)^{j_{h}+j_{\overline{\alpha}}}\left[ \frac{%
\langle\overline{\phi }_h\|P_L\|\phi _{ \overline{\alpha}
}\rangle\langle\overline{\phi }_{ \overline{\alpha}
}\|Q_L\|\phi_h\rangle}{E-(\varepsilon_{ \overline{\alpha}
}-\varepsilon_h)+i\eta }
-\frac{\langle\overline{\phi }_{ \overline{\alpha} }\|P_L\|\phi _h%
\rangle\langle\overline{\phi }_h\|Q_L\|\phi _{ \overline{\alpha} }\rangle}{E+(\varepsilon_{ \overline{\alpha} }-\varepsilon_h)+i\eta }%
\right]\right\}\ ~,  \label{eq3}
\end{eqnarray}
\end{widetext}

The unperturbed polarization operator includes not only the positive energy
particle-hole pairs but also pairs formed from the Dirac sea states and
Fermi sea states. In Refs.\cite{Ma01,Ring01} the authors show that the
inclusion of configurations built from the positive energy states in the
Fermi sea and negative energy states in the Dirac sea is essential to give
correctly a quantitative description in the excitation energies of isoscalar
giant multipole resonances as well as to ensure the current conservation and
decouple the spurious states.

The pairing correlation and coupling to the continuum are
important for exotic nuclei. A proper treatment of the resonant
continuum to pairing correlations has been recently investigated
in the Hartree-Fock (HF) Bogoliubov or the HF+BCS
approximation\cite{San00,Kru01,Gra01} and the extended RMF+BCS
approximation\cite{Cao031,San03}. It shows that the simple BCS
approximation in the resonant continuum with a proper boundary
condition works well in the description of ground state properties
even for neutron rich nuclei. We shall treat the pairing
correlation in the BCS approximation in this work and the resonant
continuum is calculated by imposing an asymptotic scattering
boundary condition.

When pairing correlations are taken into account, the elementary
excitation is a two-quasiparticle excitation, rather than a
particle-hole excitation. The unperturbed polarization operator in
the QRRPA in the response function formalism can be constructed in
a similar way:

\begin{widetext}
\vglue -0.50cm
\begin{eqnarray}
&&\Pi _0^R(P,Q;k,k^{\prime };E) \nonumber \\
&&=\frac{(4\pi)^{2}}{2L+1}\left\{\sum_{\alpha,\beta}(-1)^{j_{\alpha}+j_{\beta}}A_{\alpha\beta}\left[ \frac{%
\langle\overline{\phi}_\alpha\|P_L\|\phi
_\beta\rangle\langle\overline{\phi
}_\beta\|Q_L\|\phi_\alpha\rangle}{E-(E_\alpha+E_{\beta})+i\eta }-\frac{\langle\overline{\phi }_\beta\|P_L\|\phi _\alpha\rangle%
\langle\overline{\phi}_\alpha\|Q_L\|\phi _\beta\rangle}{E+(E_\alpha+E_{\beta})+i\eta }%
\right]\right. \nonumber \\
&&\left.~~~~~~~+\sum_{\alpha,\overline{\beta}}(-1)^{j_{\alpha}+j_{\overline{\beta}}}\upsilon_{\alpha}^{2}\left[
\frac{ \langle\overline{\phi}_\alpha\|P_L\|\phi _{\overline{\beta
}}\rangle\langle\overline{\phi}_{\overline{\beta}}\|Q_L\|\phi_h\rangle}{E-(E_\alpha+\lambda-\varepsilon_{\overline{\beta
}})+i\eta }-\frac{\langle\overline{\phi
}_{\overline{\beta}}\|P_L\|\phi _\alpha
\rangle\langle\overline{\phi}_\alpha\|Q_L\|\phi
_{\overline{\beta}}\rangle}{E+(E_\alpha+\lambda-\varepsilon_{\overline{\beta}})+i\eta
} \right]\right\}\ ~,  \label{eq4}
\end{eqnarray}
\end{widetext}

with
\begin{equation}
A_{\alpha\beta}=(u_\alpha\upsilon_\beta+(-1)^{L}\upsilon_{\alpha}u_\beta)^{2}(1+\delta_{\alpha
\beta})^{-1}~, \label{eq5}
\end{equation}
where $\upsilon^{2}_\alpha$ is the occupation probability and
$u^{2}_\alpha=1-\upsilon^{2}_\alpha$.
$E_\alpha=\sqrt{(\varepsilon_\alpha-\lambda)^{2}+\Delta^{2}}$ is
the quasiparticle energy, where $\lambda$ and $\Delta$ are the
Fermi energy and pairing correlation gap, respectively. In the BCS
approximation, the $\phi_\alpha$ is the eigenfunction of the
single particle Hamiltonian with an eigenvalue
$\varepsilon_\alpha$. In Eq.(4),  terms in the first square
bracket represent those excitations with one quasiparticle in
fully or partial occupied states and one quasiparticle in partial
occupied or unoccupied states. Terms in the second square bracket
describe all excitations between positive energy fully or partial
occupied states and negative energy states in the Dirac sea. For
unoccupied positive energy states outside the pairing active
space, their energies are $E_\beta=\varepsilon_\beta-\lambda$,
occupation probabilities $\upsilon^{2}_\beta=0$ and
$u^{2}_\beta=1$. For fully occupied positive energy states, the
quasiparticle energy and the occupation probability are
$E_\alpha=\lambda-\varepsilon_\alpha$ and $\upsilon^{2}_\alpha=1$
in Eq.(4). States in the Dirac sea are not involved in pairing
correlations. Therefore those quantities
$\upsilon_{\overline{\beta}}^{2}$ and $u_{\overline{\beta}}^{2}$
are set to be $0$ and $1$, respectively. Once the unperturbed
polarization operator in the quasiparticle scheme is built, the
QRRPA response function can be obtained by solving the
Bethe-Salpeter equation (2) as usually done in the RRPA.

\section{Numerical calculation and test of the QRRPA}

In this section, we first check the validity of the present QRRPA
calculations. We apply the QRRPA to calculate the response
function of the isovector giant dipole resonance(IVGDR) and the
isoscalar giant quadrupole resonance(ISGQR) in the neutron rich
nucleus $^{22}$O . Similar calculations for the nucleus $^{22}$O
were recently performed by Paar et al\cite{Pa03} in the framework
of the Relativistic Hartree-Bogoliubov(RHB) + QRRPA in the
configuration space formalism.

\begin{figure}[hbtp]
\includegraphics[scale=0.5]{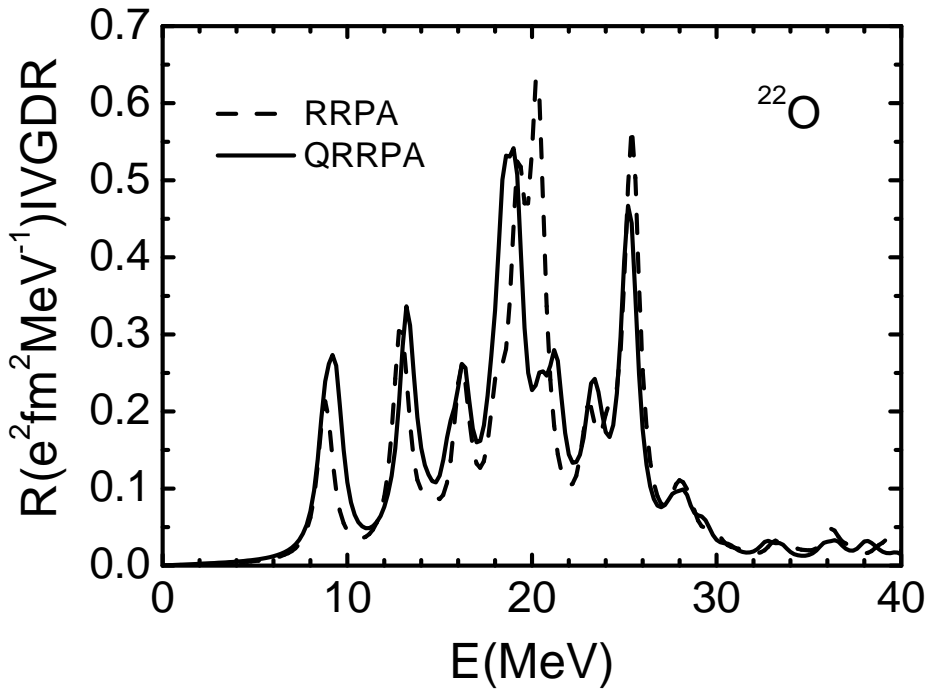}
\vglue -4.50cm \caption{ IVGDR strengths in $^{22}$O. The solid curve
represents the result calculated in the QRRPA approach. The result performed
in the RRPA approach is shown by a dashed curve. All results are calculated
with the effective Lagrangian parameter set NL3.} \label{Fig.1}
\end{figure}

The ground state properties of the nucleus $^{22}$O are calculated
in the extended RMF+BCS approach\cite{Cao031} with the parameter
set NL3. The neutron pairing gap is obtained from the experimental
binding energies of neighboring nuclei, $\Delta_n = 1.532$ MeV. In
the QRRPA calculation particle-hole residual interactions are
taken from the same effective interaction NL3, which is used in
the description of the ground state of $^{22}$O. Fully occupied
states and states in the pairing active space are calculated
self-consistently in the extended RMF+BCS approach in the
coordinate space. The BCS active space is taken as all states in
the $sd$ shell as well as 1f$_{7/2}$ state, which is a resonant
state in $^{22}$O. A scattering boundary condition is imposed in
the resonant continuum. Unoccupied states outside of the pairing
active space are obtained by solving the Dirac equation in the
expansion on a set of the harmonic oscillator basis. The response
functions of the nuclear system to the external operator are
calculated at the limit of zero momentum transfer. It is also
necessary to include the space-like parts of vector mesons in the
QRRPA calculations, although they do not play role in the ground
state\cite{Ring96}. The consistent treatment guarantees the
conservation of the vector current.

In Fig.1, we show the response function of the IVGDR mode in
$^{22}$O calculated in the RRPA and QRRPA approaches. The
isovector dipole operator used in the calculations is\cite{Liu91}:
\begin{equation}
Q=e\frac{N}{A}\sum_{i=1}^Zr_iY_{1M}(\hat{r}_i)
  -e\frac{Z}{A}\sum_{i=1}^Nr_iY_{1M}(\hat{r}_i)~,
\label{eq6}
\end{equation}
which excites an $L=1$ type electric (spin-non-flip) $\Delta T=1$
and $\Delta S=0$ giant resonance with $J^\pi =1^{-}$.  The
spurious state for exotic nuclei may appear at the energy around 1
MeV in the IVGDR strength in numerical calculations due to the
mixture of the isoscalar mode\cite{Cao01}. In our present
calculations, the spurious state is removed by slightly adjusting
the coupling constant of the $\sigma$ meson in the residual
interaction by less than 1\%, that does not affect the general
results.

In general, the IVGDR strengths in light stable nuclei are
expected to be fragmented substantially. This also occurs in the
response function of the IVGDR in neutron rich nuclei. More
fragmented distributions around the GDR region in $^{22}$O are
observed in Fig.1. In addition to the characteristic peak of the
IVGDR at the energy around 20 MeV, the low-lying dipole strength
appears at the excitation energy below 10 MeV.  It can be seen
that the inclusion of pairing correlations enhances the low-lying
dipole strength and has a slight effect on the strength at the
normal dipole resonance.  This illustrates the importance of
including pairing correlations in the study of the low-lying
isovector dipole strength in neutron rich nuclei. The effect of
pairing correlations on the isovector dipole strength in $^{22}$O
observed in our calculation is consistent with that obtained in
the RHB+QRRPA in the configuration formalism (Fig.2 of
Ref.\cite{Pa03}).

\begin{figure}[hbtp]
\includegraphics[scale=0.5]{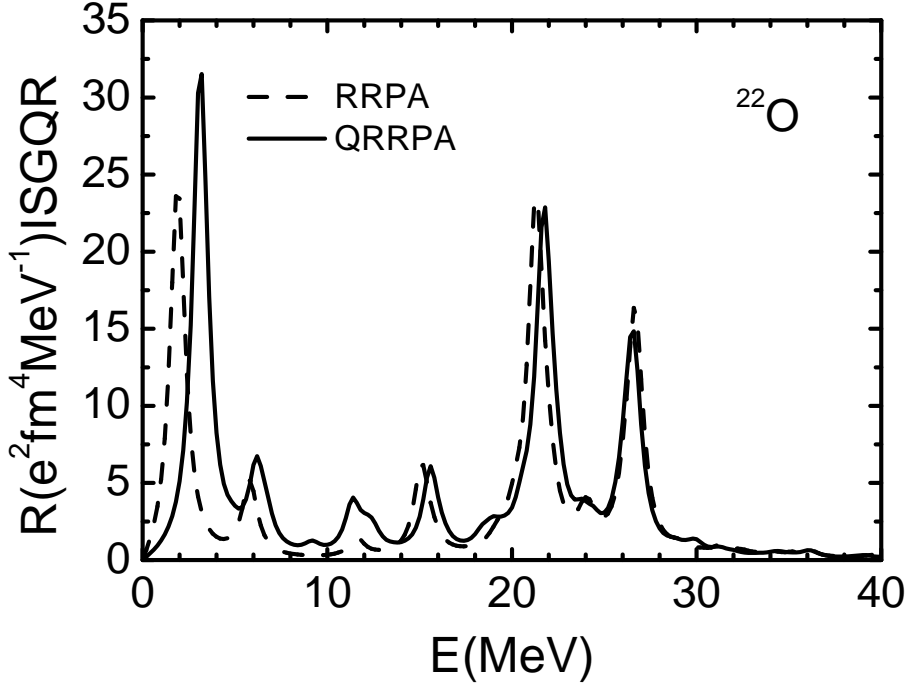}
\vglue -4.50cm \caption{ISGQR strengths in $^{22}$O. Notations are the same
as in Fig.1.} \label{Fig.2}
\end{figure}

The response functions of the ISGQR mode in $^{22}$O calculated in
the RRPA and QRRPA approaches are shown in Fig.2. In present
calculation the isoscalar quadrupole operator is taken from
Ref.\cite{Liu91}:
\begin{equation}
Q=e\frac{Z}{A}\sum_{i=1}^Ar_i^2Y_{2M}(\hat{r}_i)~, \label{eq7}
\end{equation}

The inclusion of pairing correlations shifts the low-lying
quadrupole strength to higher energy region and enhances the
low-lying quadrupole strength, while this only slightly affect the
strength at the normal giant resonance region. A similar result on
the isoscalar quadrupole strength in $^{22}$O has been observed in
Fig.3 of Ref.\cite{Pa03} in the framework of the RHB+QRRPA.

\begin{table}
\caption{ The energy weighted moment $m_1$ at $E_x<$ 60 MeV for
the electric isovector dipole and isoscalar quadrupole excitations
in $^{22}$O. DC represents the result from double commutators in
the non-relativistic approach. }
\begin{ruledtabular}
\begin{tabular}{cccc}

 &   DC     &    \multicolumn{2}{c}{$m_1(E_x<$60 MeV)} \\
 &          &    RRPA    &    QRRPA\\

\hline

IVGDR($e^2fm^2$MeV)   &   75.9   &   82.08 &   81.71   \\
ISGQR($e^2fm^4$MeV)   &   2018   &   1971  &   2159    \\

\end{tabular}
\end{ruledtabular}
\end{table}

In Table I we show the energy weighted moment $m_1$ at $E_x<$60
MeV for electric isovector dipole and isoscalar quadrupole
excitations in $^{22}$O. DC represents the result from double
commutators in the non-relativistic approach. In the isovector
dipole mode the value corresponds to the Thomas-Reiche-Kuhn (TRK)
dipole sum rule. It is shown that the $m_1$ obtained by the
integration of the RPA strength till to 60 MeV is slightly larger
than that obtained by the double commutator (DC) in the dipole
mode. While in the quadrupole mode both RRPA and QRRPA results are
close to the DC value.

\section{Isovector dipole excitation in neutron rich nuclei $^{26}$Ne and
$^{28}$Ne}

\subsection{Ground state properties of nuclei $^{26}$Ne
and $^{28}$Ne}

Ground state properties of nuclei $^{26}$Ne and $^{28}$Ne are
studied in the extended RMF+BCS with the parameter set NL3, where
a spherical symmetry is assumed. The continuum is calculated by
imposing a scattering boundary condition and the width of the
resonant state is not considered in this work. Constant pairing
gaps are adopted in the calculation of pairing correlations, which
are obtained from the experimental binding energies of neighboring
nuclei by the formula:

\begin{widetext}
\vglue -0.50cm
\begin{equation}
\Delta_{p}=\frac{1}{8}\left(B(Z-2,N)-4B(Z-1,N)+6B(Z,N)-4B(Z+1,N)+B(Z+2,N)\right)
~, \label{eq8}
\end{equation}
\begin{equation}
\Delta_{n}=\frac{1}{8}\left(B(Z,N-2)-4B(Z,N-1)+6B(Z,N)-4B(Z,N+1)+B(Z,N+2)\right)
~. \label{eq9}
\end{equation}
\end{widetext}

In our calculations, the neutron pairing active space in nuclei
$^{26}$Ne and $^{28}$Ne includes states up to the $N = 28$ major
shell and $2p_{3/2}$ state, which are  $1d_{5/2}$, $2s_{1/2}$,
$1d_{3/2}$, $2p_{3/2}$, and $1f_{7/2}$. The BCS active space for
proton is taken as all states in the $sd$ shell. In Table II we
list the neutron and proton pairing gaps in nuclei $^{26}$Ne and
$^{28}$Ne derived from Eqs.(8,9) and the calculated ground state
properties, including neutron and proton Fermi energies, total
binding energies as well as the neutron and proton rms radii. The
values in the parenthesis are corresponding experimental binding
energies taken from Ref.\cite{Aud95}. Neutron single particle
energies and BCS occupation probabilities for those states near
the neutron Fermi energy are shown in Table III, where levels
($2p_{3/2}$ and $1f_{7/2}$) with positive energies are the single
particle resonant states.

\begin{table}
\caption{Neutron and proton pairing gaps in $^{26}$Ne and
$^{28}$Ne, and calculated ground state properties: neutron and
proton Fermi energies, binding energies as well as neutron and
proton rms radii. Values in the parenthesis are the corresponding
experimental data of the binding energy\cite{Aud95}.}
\begin{ruledtabular}
\begin{tabular}{ccc}

 & $^{26}$Ne & $^{28}$Ne \\
\hline

$\Delta_n(MeV)$ & 1.436         & 1.400  \\
$\Delta_p(MeV)$ & 2.025         & 2.101  \\
$\lambda_n(MeV)$& -5.325        & -4.290 \\
$\lambda_p(MeV)$& -14.168       & -16.247 \\
$E_B(MeV)$      & 201.8(201.6)  & 210.5(206.9)  \\
$r_n(fm)$       & 3.179         & 3.348  \\
$r_p(fm)$       & 2.784         & 2.833  \\
\end{tabular}
\end{ruledtabular}
\end{table}

\begin{table}
\caption{Neutron single-particle energies $\varepsilon_{\alpha}$
and occupation probabilities $\upsilon^{2}_{\alpha}$ of levels
near the neutron Fermi energy in nuclei $^{26}$Ne and $^{28}$Ne.}
\begin{ruledtabular}
\begin{tabular}{crcrc}
 &\multicolumn{2}{c}{$^{26}$Ne}&\multicolumn{2}{c}{$^{28}$Ne}\\
 &$\varepsilon_{\alpha}$(MeV)&$\upsilon^{2}_{\alpha}$&$\varepsilon_{\alpha}$(MeV)&$\upsilon^{2}_{\alpha}$\\
\hline

$1d_{5/2}$  &-10.548  &0.982 &-10.836  &0.989  \\
$2s_{1/2}$  &-6.549   &0.824 &-7.054   &0.946  \\
$1d_{3/2}$  &-3.408   &0.099 &-4.299   &0.503  \\
$2p_{3/2}$  &         &      &0.786    &0.018  \\
$1f_{7/2}$  &2.946    &0.007 &2.223    &0.011  \\

\end{tabular}
\end{ruledtabular}
\end{table}

\begin{figure}[hbtp]
\includegraphics[scale=0.5]{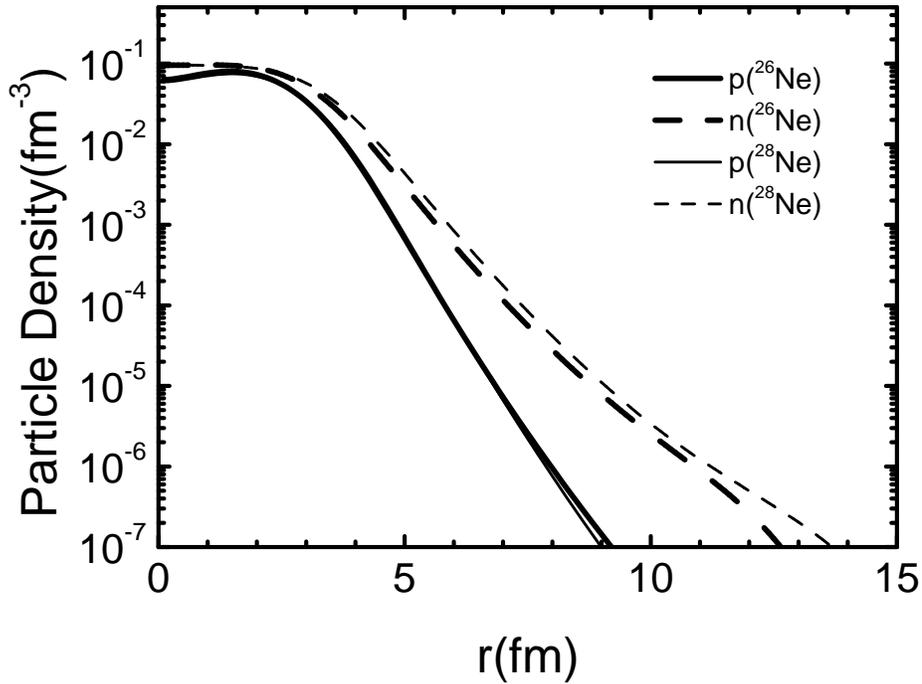}
\vglue -4.5cm \caption{Neutron and proton density distributions in nuclei
$^{26}$Ne and $^{28}$Ne. All results are calculated in the RMF+BCS.}
\label{Fig.3}
\end{figure}

Thus the nucleon density with pairing correlations
can be written as:

\begin{equation}
\rho(r)=\sum_\alpha\frac{(2j_\alpha
+1)}{4\pi}\upsilon_\alpha^{2}\phi_\alpha^{\dagger}(r)\phi_\alpha(r)
~, \label{eq10}
\end{equation}
where the summation runs over all states weighted by the factor
$\upsilon_\alpha^{2}$. In Fig.3 we show the calculated nucleon
densities in $^{26}$Ne and $^{28}$Ne. It is shown in Table II that
proton rms radii are much smaller than those of neutron.  The
neutron densities have far extending tails seen in Fig.3, which
clearly shows neutron skins formed in those nuclei.

\subsection{Low-lying isovector dipole modes in
$^{26}$Ne and $^{28}$Ne}

In this work a spherical assumption is adopted in the RMF and RRPA
calculation, although $^{26}$Ne and $^{28}$Ne are deformed and anharmonic.
In order to see how far the present model can be effective one first studies
the quadrupole excitations in these nuclei and compare with the experimental
data. The calculated energies and B(E2) values for the lowest $2^+$ states
in $^{26}$Ne and $^{28}$Ne are listed in Table IV. The values in the
parenthesis are the corresponding experimental data taken from
Ref.\cite{Pri99}. Although the B(E2) value in $^{28}$Ne is very close to the
lower limit of experimental data, the present calculations reasonably
reproduce the lowest $2^+$ states and its B(E2) values.

We now apply the QRRPA approach to investigate the isovector dipole response
in $^{26}$Ne and $^{28}$Ne. We focus our attention on properties of the
isovector low-lying dipole strength. In Fig.4 we present the Hartree and
perturbed strengths for the isovector dipole mode in nuclei $^{26}$Ne (upper
panel) and $^{28}$Ne (lower panel) calculated in the RRPA and QRRPA
approaches. Short-dashed curves represent the RRPA strengths, the QRRPA
response functions are denoted by solid ones. In Fig.4, it can be seen that
the perturbed strengths at the energy above 10 MeV for those neutron rich
nuclei are very fragmented. Compared to the Hartree strengths, the RPA
strengths are shifted to higher energy region due to the repulsive
particle-hole residual interaction generated mainly by exchanging $\rho$
meson.

\begin{table}
\caption{ The calculated energies and B(E2) values for the lowest $2^+$
states in $^{26}$Ne and $^{28}$Ne. The values in the parenthesis are the
corresponding experimental data taken from Ref.\cite{Pri99}.}
\begin{ruledtabular}
\begin{tabular}{ccc}

  &    $E_{2^{+}}$(MeV)      &  $B(E2)$($e^{2}fm^{4}$)\\
\hline
$^{26}$Ne   &   1.46(1.99$\pm$0.012)   &   223(228$\pm$41)  \\
$^{28}$Ne   &   1.29(1.32$\pm$0.020)   &   156(269$\pm$136)  \\
\end{tabular}
\end{ruledtabular}
\end{table}

In addition to the characteristic peak of the IVGDR around energy
of 20 MeV, low-lying dipole strengths appear at the excitation
energy below 10 MeV. It is shown in Fig.4 that the effect of
pairing correlations on the isovector dipole strength in $^{26}$Ne
and $^{28}$Ne shifts the low-lying dipole strength to higher
energy region and decreases the low-lying dipole strength,
especially in $^{26}$Ne. The situation slightly differs from that
in $^{22}$O, where the low-lying strength is increased. The
decrease of the low-lying strength is mainly due to the fact that
two-quasiparticle excitations in the dipole mode are weakened by a
factor of $\upsilon_\alpha^{2}$ when the pairing correlation is
switched on. In addition, the two-quasiparticle excitation energy
is larger than the corresponding particle-hole excitation energy
in the dipole mode\cite{Hag01}. On the contrary, when the pairing
correlation is taken into account the configuration space becomes
larger, which allows for the particle-particle and hole-hole
transitions. This enlarged configuration space may increase the
low-lying strength. As analyzed below, the state $2s_{1/2}$
produces a large contribution to the low-lying strength of the
dipole mode in $^{26,28}$Ne and $^{22}$O. It is known that the
$2s_{1/2}$ state is located below and above the Fermi surface in
$^{26,28}$Ne and $^{22}$O, respectively. Therefore
particle-particle excitations, especially due to the particle
state $2s_{1/2}$, largely enhance the low-lying strength in
$^{22}$O, which is not true in the case of $^{26}$Ne and
$^{28}$Ne.

\begin{figure}[hbtp]
\includegraphics[scale=0.5]{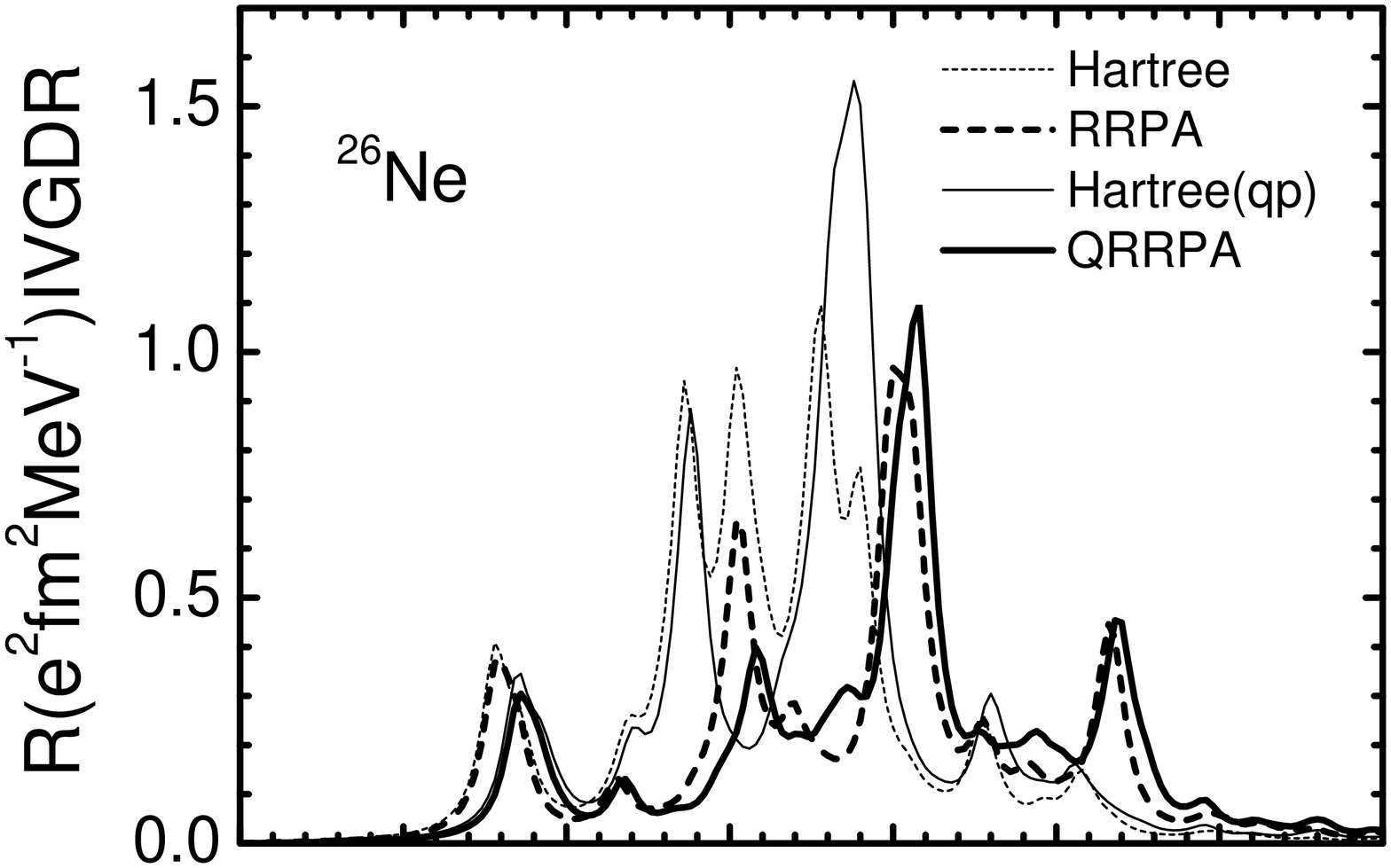}
\vglue -6.0cm
\includegraphics[scale=0.5]{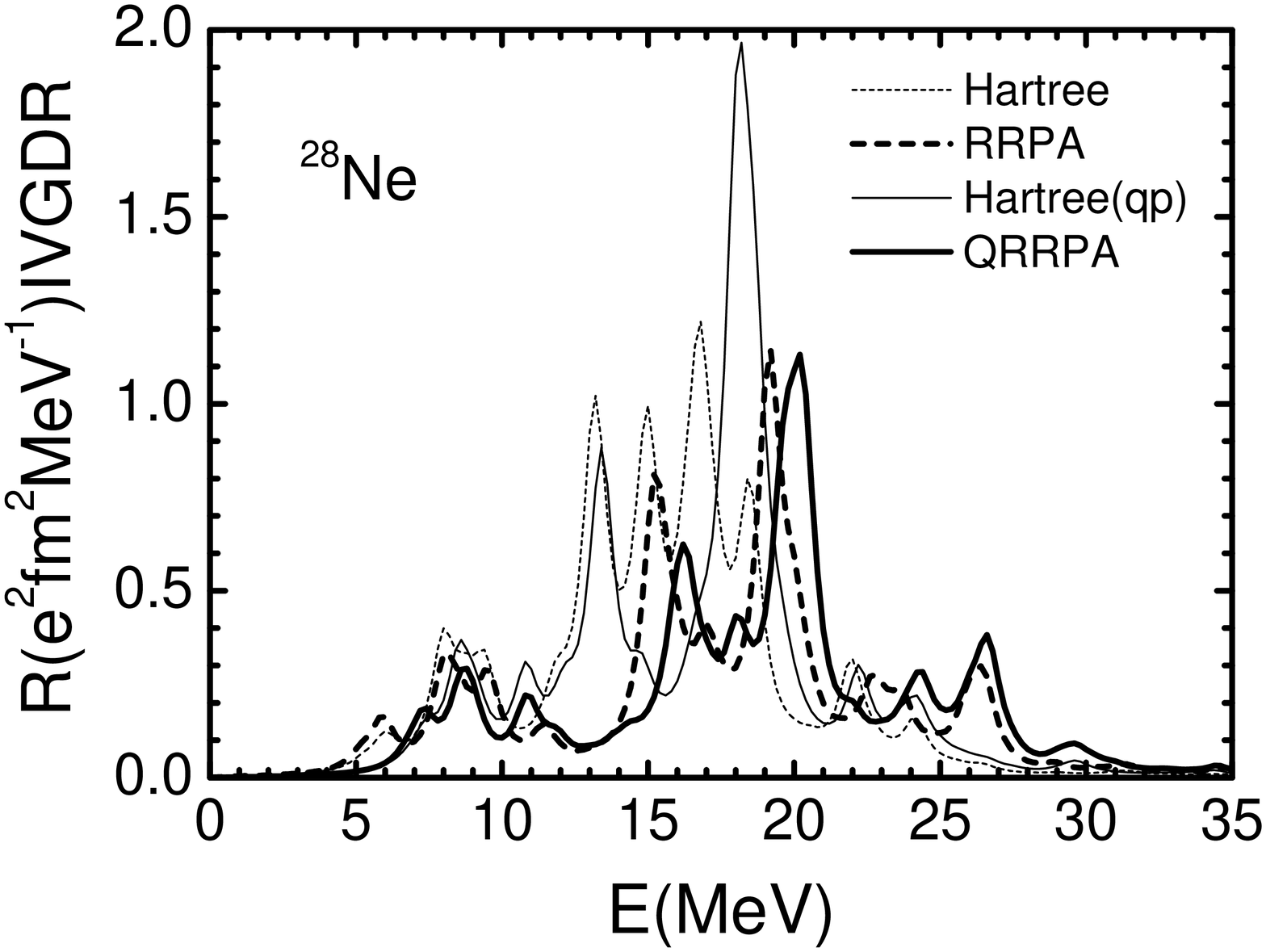}
\vglue -4.5cm \caption{Isovector dipole strength functions in neutron rich
nuclei $^{26}$Ne (upper panel) and $^{28}$Ne (lower panel). The QRRPA
responses with the pairing (solid curves) are compared with the RRPA
calculation without the pairing (short-dashed curves).  The thick and thin
curves represent perturbed and Hartree strengths, respectively.  }
\label{Fig.4}
\end{figure}

\begin{figure}[hbtp]
\includegraphics[scale=0.5]{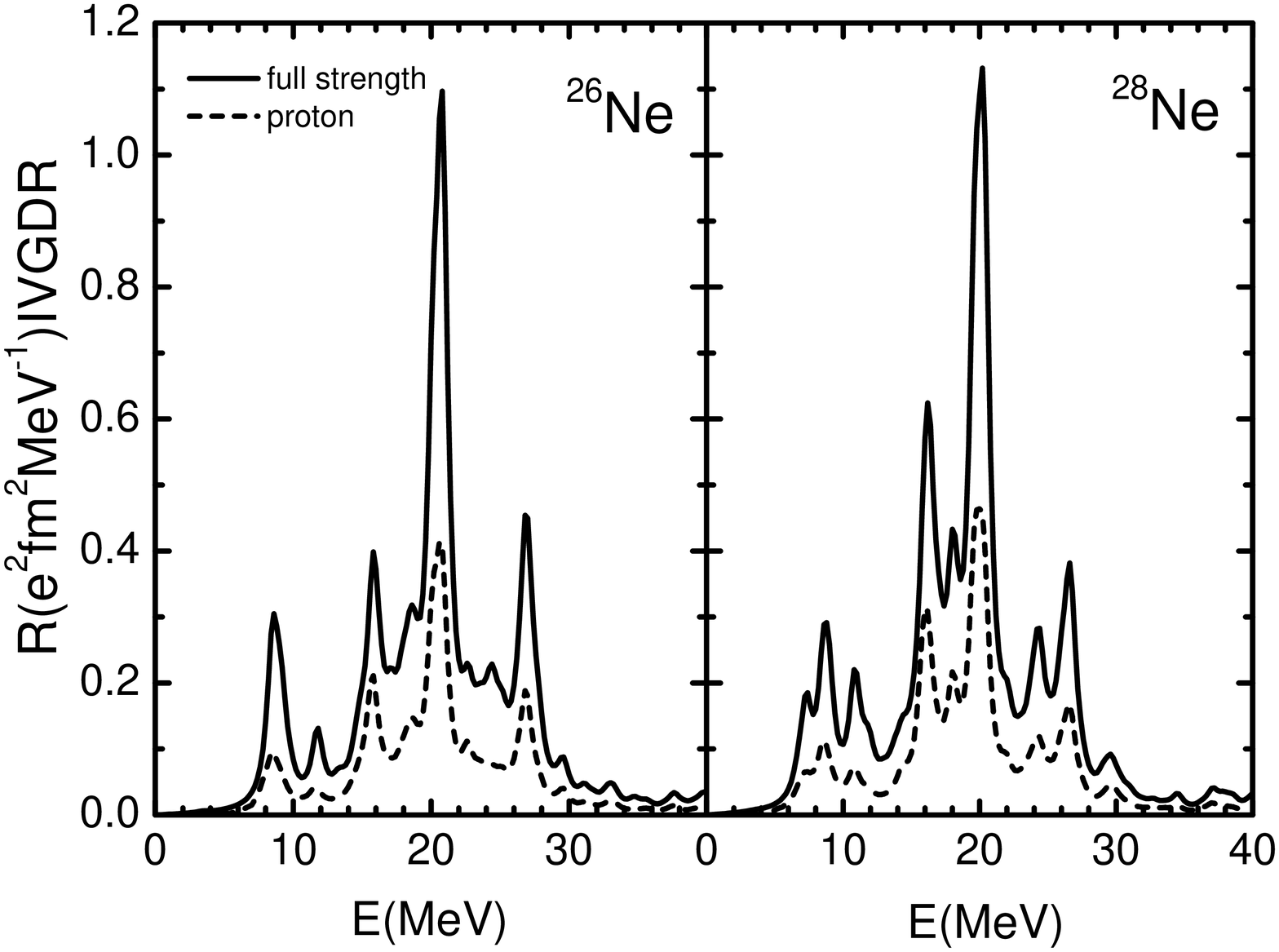}
\vglue -3.50cm \caption{The contribution of proton to the isovector dipole
strength in $^{26}$Ne (left panel) and $^{28}$Ne (right panel) in the QRRPA
approach. The solid curves are the full QRRPA strengths. The short-dashed
curves represent the strengths from proton.} \label{Fig.5}
\end{figure}

 In comparison with the Hartree strength it is
found that the RPA strength of the isovector dipole mode at the
low energy region in $^{26}$Ne and $^{28}$Ne shown in Fig.4
remains at its position and is contributed mainly from a few
particle-hole configurations, which shows a single particle like
property. The low-lying strength is slightly attracted back to the
lower energy, which is due to the correlations of the isoscalar
operator in the isovector mode\cite{Cao01}. Differing from the
normal IVGDR response, the low-lying resonance can be interpreted
as the excitation of the excess neutrons out of phase with the
core formed with an equal number of protons and
neutrons\cite{Vre01}. Analyzing the Hartree strength of the
isovector dipole mode in $^{26}$Ne at the energy below 10 MeV
calculated with pairing correlations, one finds a pronounced peak
around 8.5 MeV, which is formed mainly from the neutron
configurations of $\nu(2s_{1/2}^{-1} 2p_{3/2})$(8.482 MeV) and
$\nu(2s_{1/2}^{-1}2p_{1/2})$ (9.232 MeV), where the value in the
parenthesis is its Hartree energy.  Since 16 neutrons in $^{26}$Ne
fill neutron orbits up to $2s_{1/2}$ and form a sub-closed shell,
the occupation probabilities at $1d_{3/2}$ and $1f_{7/2}$ states
in $^{26}$Ne are relatively small, see Table III. Therefore the
contribution from neutron states $1d_{3/2}$ and $1f_{7/2}$ to the
low-lying Hartree strength is insignificant. In contrast the
Hartree strength at the low energy region in $^{28}$Ne, in
addition to the peaks formed from the neutron configurations of
$\nu(2s_{1/2}^{-1}2p_{3/2})$(8.523 MeV) and
$\nu(2s_{1/2}^{-1}2p_{1/2})$ (9.080 MeV), a few more peaks appear,
which are formed from the neutron excitation between the bound
level $1d_{3/2}$ and levels in the continuum. It is found that the
Hartree strengths at the low-lying dipole in $^{26}$Ne and
$^{28}$Ne are mainly due to neutron excitations near the Fermi
surface.

\begin{table}
\caption{The non-energy weighted moment $m_0$ and energy weighted moment
$m_1$ of isovector dipole strengths  in $^{26}$Ne and $^{28}$Ne in the QRRPA
calculations. We separate the energy region into the low energy (0 MeV $\leq
E_x \leq$ 10 MeV) and the high energy (10 MeV $\leq E_x \leq$ 30 MeV). The
values in the last two column are obtained from the classical TRK dipole sum
rule and TRK cluster sum rule($e^2fm^2$MeV). The units are $e^2fm^2$ and
$e^2fm^2$MeV for $m_0$ and $m_1$, respectively.}
\begin{ruledtabular}
\begin{tabular}{ccccccc}

 &\multicolumn{2}{c}{0MeV$\leq E_x \leq$10MeV}&\multicolumn{2}{c}{10MeV$\leq E_x \leq$30MeV}& \\
 &$m_{0}$ &$m_{1}$ &$m_{0}$ &$m_{1}$ &  S$_{\text{TRK}}$ & S$_{\text{Clus}}$   \\

\hline

 $^{26}$Ne  & 0.542  & 4.525 & 5.032  & 103.9  & 91.7 & 17.2 \\
 $^{28}$Ne  & 0.705  & 5.606 & 5.648  & 111.2  & 95.8 & 21.3 \\

\end{tabular}
\end{ruledtabular}
\end{table}

\begin{table}
\caption{Centroid energies of the isovector dipole response functions in
$^{26}$Ne and $^{28}$Ne. The centroid energies are calculated within 0
$\sim$ 10 MeV and 10 $\sim$ 60 MeV, respectively. All energy values are in
unit of MeV. }
\begin{ruledtabular}
\begin{tabular}{ccc}
    &   $^{26}$Ne    &    $^{28}$Ne \\
\hline

$\overline{E}$(0 $\sim$ 10)  & 8.34       & 7.94     \\
$\overline{E}$(10 $\sim$ 60) & 22.32      & 21.13    \\

\end{tabular}
\end{ruledtabular}
\end{table}

Although the Hartree low-lying strength is mainly formed from the neutron
excitations, the RPA strengths are fully correlated and contributed from
both neutron and proton. To illustrate the contribution of the proton to the
full strength we set a very small value of the neutron effective charge in
the dipole operator instead of $eZ/A$. Results are plotted in Fig.5. The
short-dashed curves represent the strengths from proton, which are compared
with the full QRRPA strengths denoted by solid curves. It clearly shows that
the proton also plays an important role in the perturbed strength even at
low-lying dipole states.

In Ref.\cite{Vre01}, the authors have studied the evolution of collectivity
in the isovector dipole response in the low-lying region for neutron-rich
isotopes of O, Ca, Ni, Zr, and Sn using RRPA method. They conclude that in
light neutron-rich nuclei, such as neutron-rich isotopes of O and Ca, the
onset of dipole strength in the low-lying region is due to single-particle
excitations of the loosely bound neutrons. By analyzing the structure of
RRPA strengths in low-lying region in nuclei $^{26}$Ne and $^{28}$Ne, we
find that there are only several configurations contributing to the
low-lying RRPA strengths and the RRPA strengths remain their positions
compared to the Hartree strengths in the low-energy region, which mean that
the RRPA strengths in low-lying region in nuclei $^{26}$Ne and $^{28}$Ne are
dominated by single-particle transitions.

In order to give a more clear description of those isovector
low-lying dipole states obtained in the QRRPA approach, we
calculate various moments of isovector dipole strengths at a given
energy interval:
\begin{equation}
m_{k}=\int_{E_1}^{E_{2}}R^{L}(E)E^{k}dE~, \label{eq11}
\end{equation}
where $E_{1}$ and $E_{2}$ are the lower and upper energies of  the integral,
respectively. The RPA equation is solved till $E = 60$ MeV in the present
calculations.  The non-energy weighted moment $m_0$ and the energy weighted
moment $m_1$ of isovector dipole strengths in $^{26}$Ne and $^{28}$Ne are
calculated at two energy intervals: the low energy region (0 MeV $\leq E_x
\leq$ 10 MeV) and the high energy region (10 MeV $\leq E_x \leq$ 30 MeV),
which are listed in Table IV. The values obtained from the
Thomas-Reiche-Kuhn (TRK) dipole sum rule are listed in the sixth column. In
our present QRRPA calculations the low-lying isovector dipole strengths in
$^{26}$Ne and $^{28}$Ne exhaust about 4.93\% and 5.85\% of the TRK dipole
sum rule, respectively. This is consistent with the recent experimental
observations in $^{18}$O, $^{20}$O and $^{22}$O, where the low-lying
isovector dipole strengths exhaust about 5\% of the TRK dipole sum
rule\cite{Aum96}. In general the percentage of the low-lying isovector
dipole strength becomes large as the increase of the neutron
excess\cite{Vre01}.  The energy weighted moment $m_1$ at $E_x<10$ MeV in
$^{28}$Ne is about 1.0\% larger than that in $^{26}$Ne.

On the other hand, the cluster sum rule\cite{Lei01,Aum99,Alh82,Hen04} is
usually used to understand the properties of low-lying dipole strength in
exotic nucleus excitations. Here we choose $^{20}$Ne as the core. In Table V
we also list the values obtained form TRK cluster dipole sum rule, say 17.2
$e^2fm^2$MeV and 21.3 $e^2fm^2$MeV for $^{26}$Ne and $^{28}$Ne,
respectively. It shows that the low-lying dipole excitations in these two
neutron rich nuclei exhaust about 26.3\% of the TRK cluster dipole sum rule.
Similar results are obtained in Ref.\cite{Lei01} for neutron rich Oxygen
isotopes.

The centroid energy of the response function is defined as:
\begin{equation}
\overline{E}=m_{1}/m_{0}~. \label{eq12}
\end{equation}
We separate the energy interval into two regions: 0 MeV$<E_x<$10
MeV and 10 MeV$<E_x<$60 MeV. The centroid energies in these two
energy regions are listed in Table V.  In the QRRPA calculations
the centroid energies of low-lying isovector dipole strengths in
$^{26}$Ne and $^{28}$Ne are 8.34 MeV and 7.94 MeV, respectively.
Whereas the centroid energies of the normal IVGDR strengths are
located at 22.3 MeV for $^{26}$Ne and 21.1 MeV for $^{28}$Ne.

\section{summary}

In this paper we have studied the properties of low-lying isovector dipole
resonances in neutron rich nuclei $^{26}$Ne and $^{28}$Ne in the framework
of the QRRPA with the effective Lagrangian parameter set NL3.  The ground
state properties are calculated in the extended RMF+BCS approach, where the
resonant continuum is properly treated. Constant pairing gaps extracted from
the experimental binding energies of neighboring nuclei are adopted in the
BCS calculation. In the QRRPA calculation the negative energy states in the
Dirac sea are included due to the completeness. It is shown that the
inclusion of pairing correlations has a relatively strong effect on the
low-lying isovector dipole strength in neutron rich nuclei. In the QRRPA
calculation the low-lying isovector dipole strengths in $^{26}$Ne and
$^{28}$Ne exhaust about 5\% and 26.3\% of the TRK dipole sum rule and TRK
dipole cluster sum rule, respectively. The centroid energies of the low-lying
dipole excitation in nuclei $^{26}$Ne and $^{28}$Ne are located at the
energy around 8.0 MeV.

\begin{acknowledgments}
We thank Professor Didier Beaumel, Nguyen Van Giai, Zong-ye Zhang
and You-wen Yu for many useful discussions. This work is supported
by the National Natural Science Foundation of China under Grant
Nos 10305014, 90103020 and 10275094, and Major State Basic
Research Development Programme in China under Contract No
G2000077400.
\end{acknowledgments}

\end{document}